# Optical and Judd–Ofelt spectroscopic investigation of $Tm^{3+}$-doped germanium–magnesium–telluroborate glasses

A. A. El-Maaref[1,*], Kh. S. Shaaban[2], E. A. Abdel Wahab[3,*]

[1] Department of Physics, College of Science, Jouf University, Sakaka, Aljouf 72341, Saudi Arabia

[2] Department of Chemistry, Faculty of Science, Al-Azhar University, P.O. 71542, Assiut, Egypt

[3] Physics Department, Faculty of Science, Al-Azhar University, Assiut, P.O 71524, Egypt

**Abstract:**

Germanium-magnesium–telluroborate glasses with the composition $78B_2O_3$–$10GeO_2$–$5TeO_2$–$(7-x)MgO$–$xTm_2O_3$ (x = 0, 0.25, 0.5, 1, and 1.5 mol%) were synthesized using the melt-quenching technique. The effect of $Tm_2O_3$ concentration on the physical, structural, and optical properties of the glasses was systematically investigated. With increasing $Tm_2O_3$ content, the density increased from 3.574 to 4.153 $g \cdot cm^{-3}$, while the molar volume decreased from 21.145 to 19.445 $cm^3 \cdot mol^{-1}$, indicating progressive structural compaction. Fourier-transform infrared (FTIR) spectra confirm the presence of $BO_3$ and $BO_4$ structural units and reveal a gradual conversion of $BO_3$ to $BO_4$ units with $Tm_2O_3$ addition, leading to an increase in bridging oxygen content and enhanced glass network connectivity. Optical properties were examined using UV–Vis absorption spectroscopy. The optical band gap decreases with increasing $Tm_2O_3$ concentration, whereas the refractive index shows a corresponding increase. The optical band gap and refractive index were found to vary within the ranges of 3.16–3.34 eV and 2.31–2.37, respectively. Several optical and physical parameters, including optical conductivity, metallization criterion, electronegativity, reflection loss, steepness parameter, and transmission coefficient, were derived and analyzed. Judd–Ofelt theory was applied to evaluate the optical intensity parameters and line strengths of $Tm^{3+}$ ions in the glass matrix. Radiative lifetimes and branching ratios for selected $Tm^{3+}$ manifolds were also calculated. The obtained results suggest that these $Tm^{3+}$-doped germanium magnesium–telluroborate glasses are promising candidates for photonic and optoelectronic applications, such as solid-state lasers, optical amplifiers, and light-emitting devices.

* Corresponding authors address: aauelmaaref@ju.edu.sa & essam.ah77@gmail.com

## 1. Introduction

Oxide glasses are amorphous materials distinguished by their high optical transparency and their ability to accommodate heavy metal oxides and lanthanide ions, such as tungsten, bismuth, erbium, and thulium, which enhance radiation absorption and optical functionality [1–3]. Owing to these features, oxide glasses have attracted considerable attention for applications requiring tailored structural and optical properties. Common glass formers such as $B_2O_3$, $SiO_2$, $TeO_2$, $GeO_2$, and $P_2O_5$ serve as suitable host matrices for rare-earth ions (REIs) because of their structural flexibility and chemical stability [2]. The properties of these network-forming oxides can be further modified through the incorporation of alkali and alkaline-earth oxides.

Among alkaline-earth modifiers, MgO plays an important role in altering the glass network structure by compensating for charge imbalances associated with non-bridging oxygens (NBOs) and bridging oxygens (BOs). In germanium–telluroborate glass systems, $Mg^{2+}$ ions exhibit strong interactions with NBOs in structural units such as [$BO_3$], [$GeO_3$], and [$TeO_3$], as well as with BOs in tetrahedral units including [$BO_4$], [$GeO_4$], and [$TeO_4$], thereby contributing to network stabilization and structural compaction [4].

Germanium magnesium–telluroborate glasses are of particular interest for photonic and nonlinear optical applications due to their favorable thermal stability, chemical durability, and distinctive optical characteristics. Recently, these glasses have been explored for various optical device applications, where rare-earth ions are frequently introduced as modifiers to enhance structural, optical, and absorption properties. Such modifications make these glass systems promising candidates for advanced photonic technologies.

Thulium oxide ($Tm_2O_3$) is a technologically significant rare-earth dopant owing to its rich energy-level structure, which enables both visible and near-infrared optical transitions. In germanium magnesium–telluroborate glass matrices, $Tm^{3+}$ ions function as active centers for upconversion and spectroscopic applications. The incorporation of $TeO_2$ is particularly advantageous, as its low phonon energy reduces non-radiative relaxation processes and enhances the emission efficiency of $Tm^{3+}$ ions, especially near the ~2 μm region, as reported in previous studies [4,5]. Additionally, investigations on $Tm^{3+}/Ho^{3+}$ co-doped tellurium-based glasses have demonstrated their suitability for laser and fiber-optic applications [6]. These findings indicate that $Tm^{3+}$-doped germanium magnesium–telluroborate glasses may serve as promising materials for photonic devices and optical fiber technologies.

The spectroscopic properties of $Tm^{3+}$ ions in both amorphous and crystalline hosts have been widely studied. For example, Dan et al. [7] reported enhanced red upconversion emission and energy transfer mechanisms in $Tm^{3+}$/Cr/Yb co-doped fluorosilicate glass-ceramics, employing time-resolved spectroscopy to evaluate radiative lifetimes. Upconversion processes and radiative parameters have also been examined in $Tm^{3+}$-doped ZBLAN glasses [8], while the luminescence behavior of $Tm^{3+}$-doped borosilicate glass-ceramics has been investigated by Santiago de la Rosa et al. [9].

In this context, it is essential to systematically examine the physical, structural, and optical properties of the present glass system. The current study focuses on elucidating the effects of $Tm_2O_3$ incorporation on chemical bonding, refractive index, optical band gap, and related optical parameters. Furthermore, spectroscopic characteristics such as line strengths and radiative lifetimes are evaluated using Judd–Ofelt theory to provide insight into the optical behavior of $Tm^{3+}$ ions in germanium magnesium–telluroborate glass matrices.

The empirical relations used for estimating optical parameters are well-established in the literature and are employed here as analytical tools to elucidate compositional trends rather than to propose new physical models.

## 2. Experimental procedure

Glasses with the chemical composition $78B_2O_3$–$10GeO_2$–$5TeO_2$–$(7-x)MgO$–$xTm_2O_3$, where x = 0, 0.25, 0.5, 1, and 1.5 mol%, were fabricated using the melt-quenching technique. The nominal compositions of the investigated samples are listed in Table 1. Analytical-grade raw materials with purity higher than 99%—$H_3BO_3$, $GeO_2$, $TeO_2$, MgO, and $Tm_2O_3$—were weighed in appropriate proportions to obtain a total batch weight of 15 g. The weighing accuracy was ±0.1 mg. The prepared mixtures were thoroughly mixed and placed in ceramic crucibles, then melted in an electrically heated furnace at 1200 °C for 30 min to ensure homogeneity. The molten glass was subsequently poured onto a stainless-steel plate and rapidly quenched at ambient conditions. The obtained glass samples were annealed at 380 °C for 4 h to relieve internal stresses and prevent crack formation. The samples were labeled as TGMB-Tm0, TGMB-Tm0.25, TGMB-Tm0.5, TGMB-Tm1, and TGMB-Tm1.5, according to the $Tm_2O_3$ content.

**Table 1.** Chemical composition of the investigated glass samples (mol%)

| Sample code | Composition (mol%) |
|---|---|
| TGMB-Tm0 | $78B_2O_3–10GeO_2–5TeO_2–7MgO$ |
| TGMB-Tm0.25 | $78B_2O_3–10GeO_2–5TeO_2–6.75MgO–0.25Tm_2O_3$ |
| TGMB-Tm0.5 | $78B_2O_3–10GeO_2–5TeO_2–6.5MgO–0.5Tm_2O_3$ |
| TGMB-Tm1 | $78B_2O_3–10GeO_2–5TeO_2–6MgO–1Tm_2O_3$ |
| TGMB-Tm1.5 | $78B_2O_3–10GeO_2–5TeO_2–5.5MgO–1.5Tm_2O_3$ |

FTIR spectra of the TGMB-Tm glass samples were recorded at room temperature using a single-beam FTIR spectrometer (JASCO FT/IR-430, Japan) in the range 400–4000 $cm^{-1}$. The glass samples were ground into fine powders and mixed with KBr at a mass ratio of 1:100 (glass:KBr). The mixture was pressed into transparent pellets under a pressure of 5 ton·$cm^{-2}$. The normalized FTIR spectra was deconvoluted into Gaussian components using Origin software to analyze the vibrational bands.

The density (ρ) of the glass samples was determined using Archimedes' method, employing toluene (density $\rho_0$ = 0.86 g·$cm^{-3}$) as the immersion liquid at room temperature. A sensitive digital balance with an accuracy of $\pm 10^{-4}$ g was used for the measurements, and the estimated random error in density determination was ±0.001 g·$cm^{-3}$. The density was calculated using:

$$\rho = \rho_0 \left(\frac{W_a}{W_a - W_t}\right) \quad (1)$$

where $W_a$ is the weight of the sample in air and $W_t$ is the weight of the sample in toluene. The molar volume ($V_m$) of the glass samples was calculated using:

$$V_m = M/\rho \quad (2)$$

where M is the molar mass of the glass, calculated as $M = \sum x_i M_i$ with $xi$ being the mole fraction and $Mi$ the molecular weight of each oxide component.

Optical absorption and transmission spectra were measured using a double-beam UV–Vis–NIR spectrophotometer (JASCO V-670, Japan) in the wavelength range of 200–2500 nm under ambient conditions.

**3. Results and Discussion:**

**3.1. Physical examinations**

The density (ρ) and molar volume ($V_m$) of the TGMB–Tm glass samples are presented in Fig. 1. The density values increase systematically with increasing $Tm_2O_3$ concentration, from 3.574 g/cm$^3$ for TGMB-Tm0 to 3.686, 3.824, 3.937, and 4.153 g/cm$^3$ for TGMB-Tm0.25, TGMB-Tm0.5, TGMB-Tm1, and TGMB-Tm1.5, respectively.

The observed increase in density can be attributed to the partial substitution of MgO, which has a relatively lower molar mass and density (40.304 g/mol and 3.6 g/cm$^3$), by $Tm_2O_3$, characterized by a much higher molar mass and density (385.867 g/mol and 8.6 g/cm$^3$) [8,9]. As the $Tm_2O_3$ content increases, the glass matrix becomes progressively denser due to the incorporation of heavier ions, leading to a reduction in free volume and structural vacancies. This behavior suggests the formation of a more compact glass network.

The corresponding molar volume ($V_m$) values show an opposite trend, decreasing from 21.145 cm$^3$/mol for TGMB-Tm0 to 20.737, 20.215, 20.074, and 19.445 cm$^3$/mol for TGMB-Tm0.25, TGMB-Tm0.5, TGMB-Tm1, and TGMB-Tm1.5, respectively. The reduction in molar volume indicates increased packing efficiency within the glass structure [4,5]. This decrease is associated with the contraction of interstitial voids and gaps within the glass network, which is consistent with the observed increase in density.

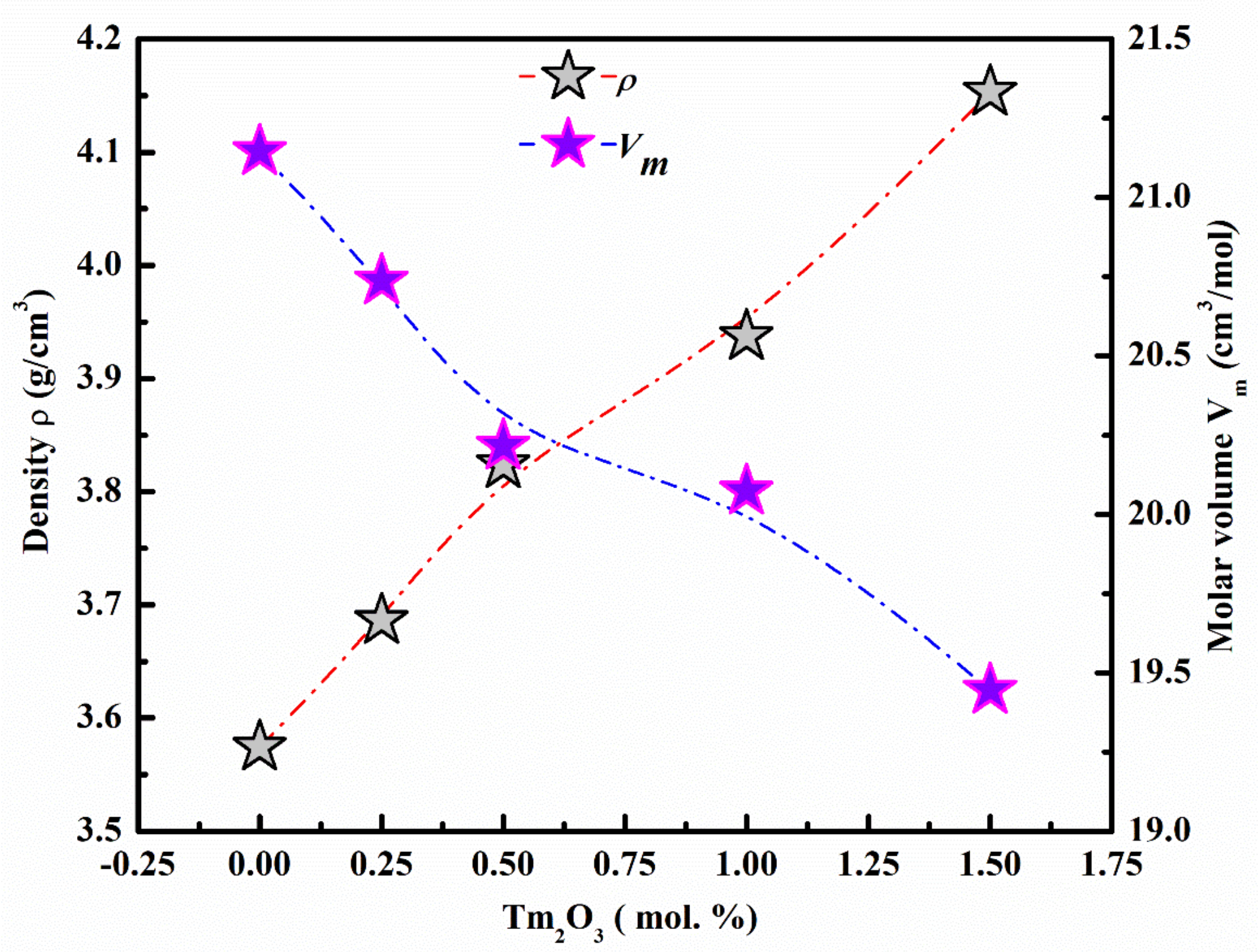

**Fig.1:** The ($\rho$) and ($V_m$) of TGMB-Tm glass system

## 3.2. FT-IR investigations

FT-IR spectroscopy is a widely used technique for investigating the structural features of glass systems and the possible interactions among their constituent units [6–16]. Figure 2 presents the FT-IR spectra of the TGMB-Tm glass system. A broad absorption band observed around ~3458 $cm^{-1}$ is attributed to the presence of absorbed $H_2O$, while the band appearing near ~2932 $cm^{-1}$ corresponds to O–H stretching vibrations.

The FT-IR spectra of the TGMB-Tm glasses reveal the presence of characteristic structural units such as $BO_3$, $BO_4$, $GeO_4$, and $TeO_4$ groups, which are commonly observed in borate-based glass systems [3]. Figure 3 illustrates the changes in the FT-IR spectra induced by varying $Tm_2O_3$ concentrations in the prepared glass network. The deconvolution parameters of the TGMB-Tm glass system are summarized in Table 2.

As shown in Table 2, the absorption bands are identified at approximately ~606, ~720, ~860, ~940, ~1070, ~1215, and ~1511 $cm^{-1}$. The assignments of these bands are as follows:
The band near ~606 $cm^{-1}$ is attributed to bending vibrations of the fundamental structural units of the glass network and is associated with asymmetric stretching vibrations of Te=O linkages in $TeO_4$ units and $TmO_6$ groups. The band observed in the range ~721–717 $cm^{-1}$ corresponds to the asymmetric stretching vibrations of Ge=O linkages in $GeO_4$ tetrahedra. The band around ~860–858 $cm^{-1}$ is related to the stretching vibrations of B–O bonds in $BO_4$ units from different borate groups.

The bands appearing in the range ~1073–941 $cm^{-1}$ are assigned to stretching vibrations of B–O bonds in diborate groups associated with $BO_4$ units. The bands located at ~1213–1223 $cm^{-1}$ are attributed to asymmetric stretching vibrations of B–O bonds in $BO_3$ units, while those in the range ~1397–1322 $cm^{-1}$ correspond to stretching vibrations of triangular $BO_3$ units. The band observed near ~1513 $cm^{-1}$ is associated with the stretching vibrations of trigonal B–O bonds in $BO_3$ units [9–19].

The fraction of tetrahedral boron units ($N_4$) in the TGMB-Tm glass system was estimated using the following relations:

$$N_4 = A_4 / (A_4 + A_3) \qquad (3)$$

$$N_3 = 1 - N_4 \qquad (4)$$

where $A_4$ represents the integrated area of the $BO_4$ bands and $A_3$ corresponds to the integrated area of the $BO_3$ bands obtained from the deconvoluted FT-IR spectra.

As shown in Fig. 4, the value of $N_4$ increases with increasing $Tm_2O_3$ concentration, while the fraction of $BO_3$ units ($N_3$) decreases. This trend indicates a gradual conversion of $BO_3$ units into $BO_4$ units, leading to an increase in bridging oxygen content, which contributes to the strengthening and stabilization of the glass network. The FT-IR results therefore support the observed trends in density and molar volume, confirming the formation of a more compact glass structure with increasing $Tm_2O_3$ content [9–19].

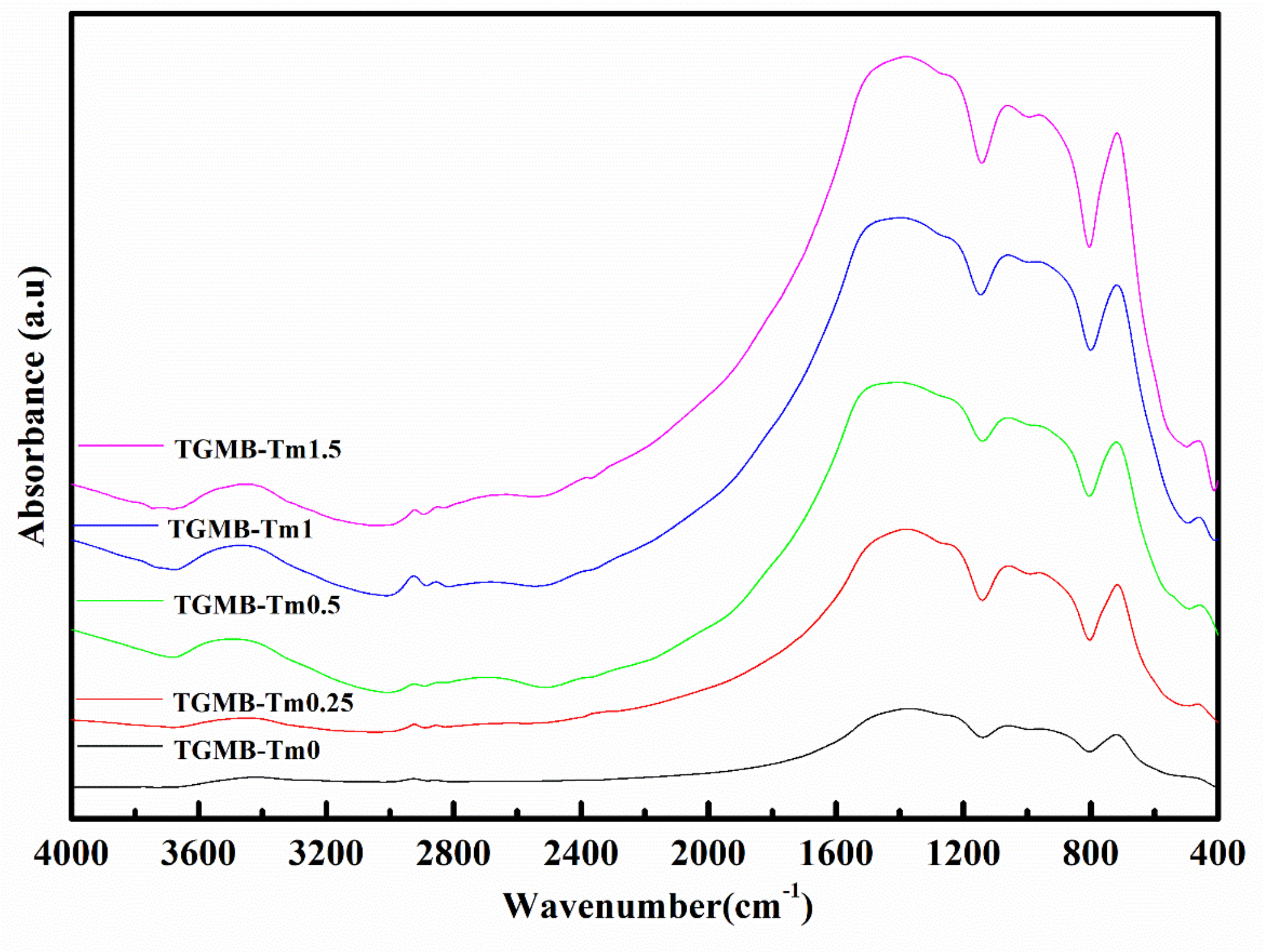

**Fig. 2:** FTIR spectra for TGMB-Tm glasses.

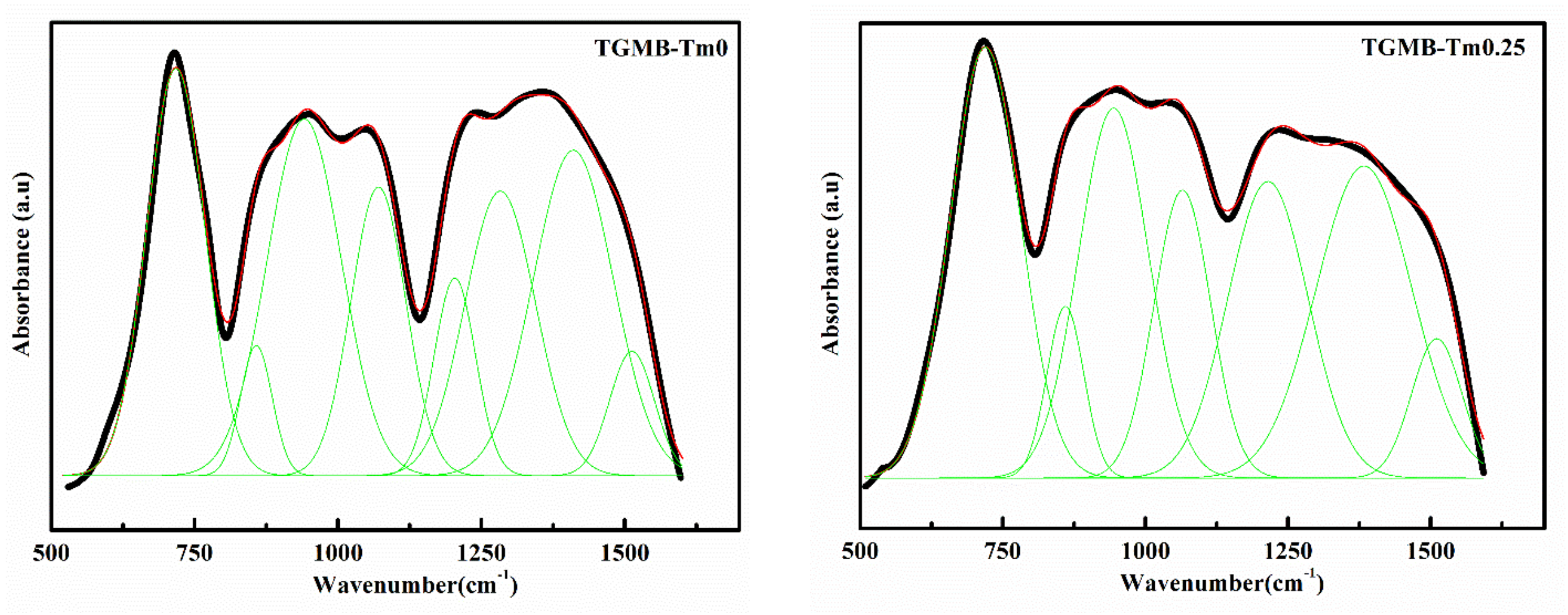

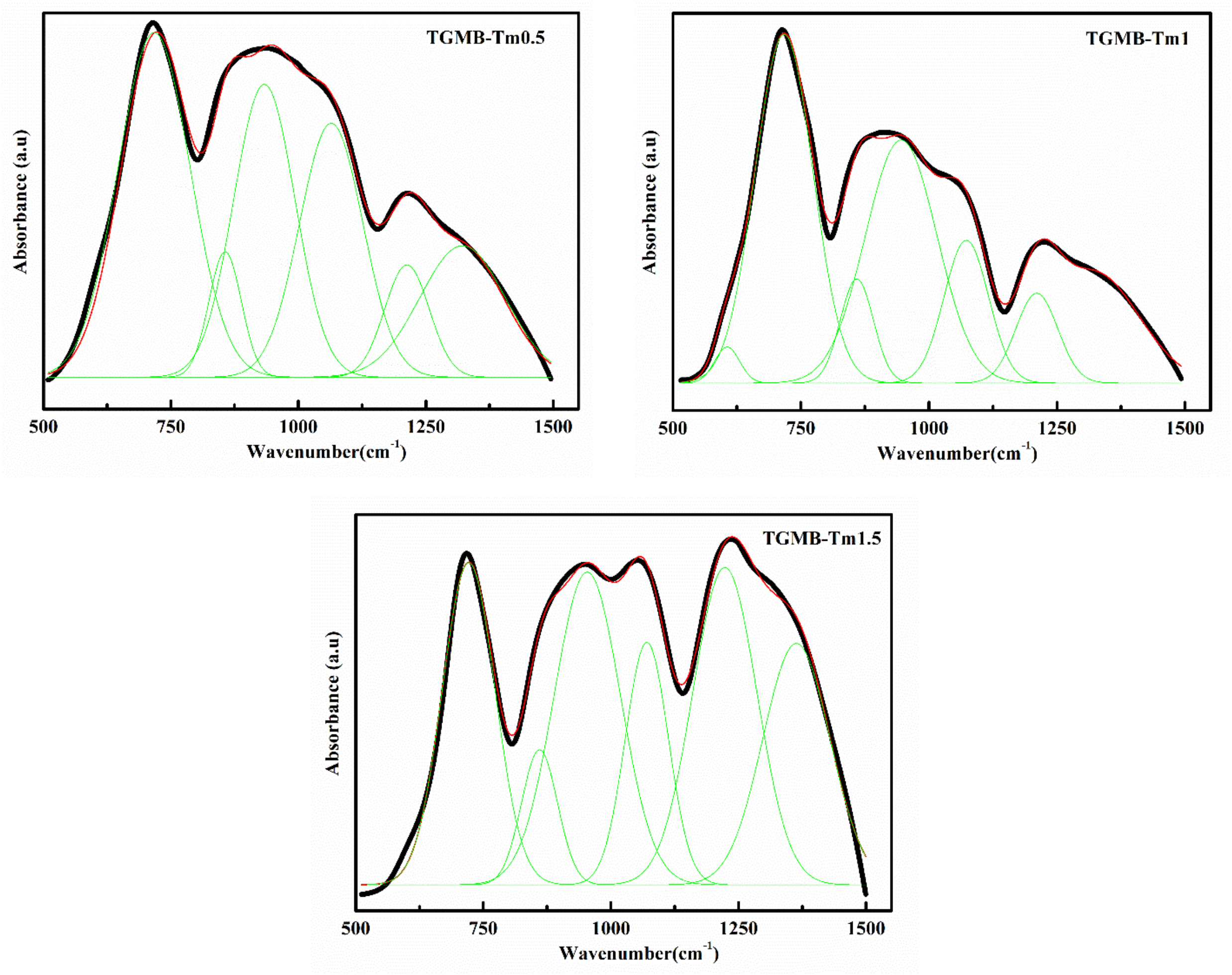

**Fig. 3:** Gaussian peaks fitting for FT-IR bands of TGMB-Tm glasses.

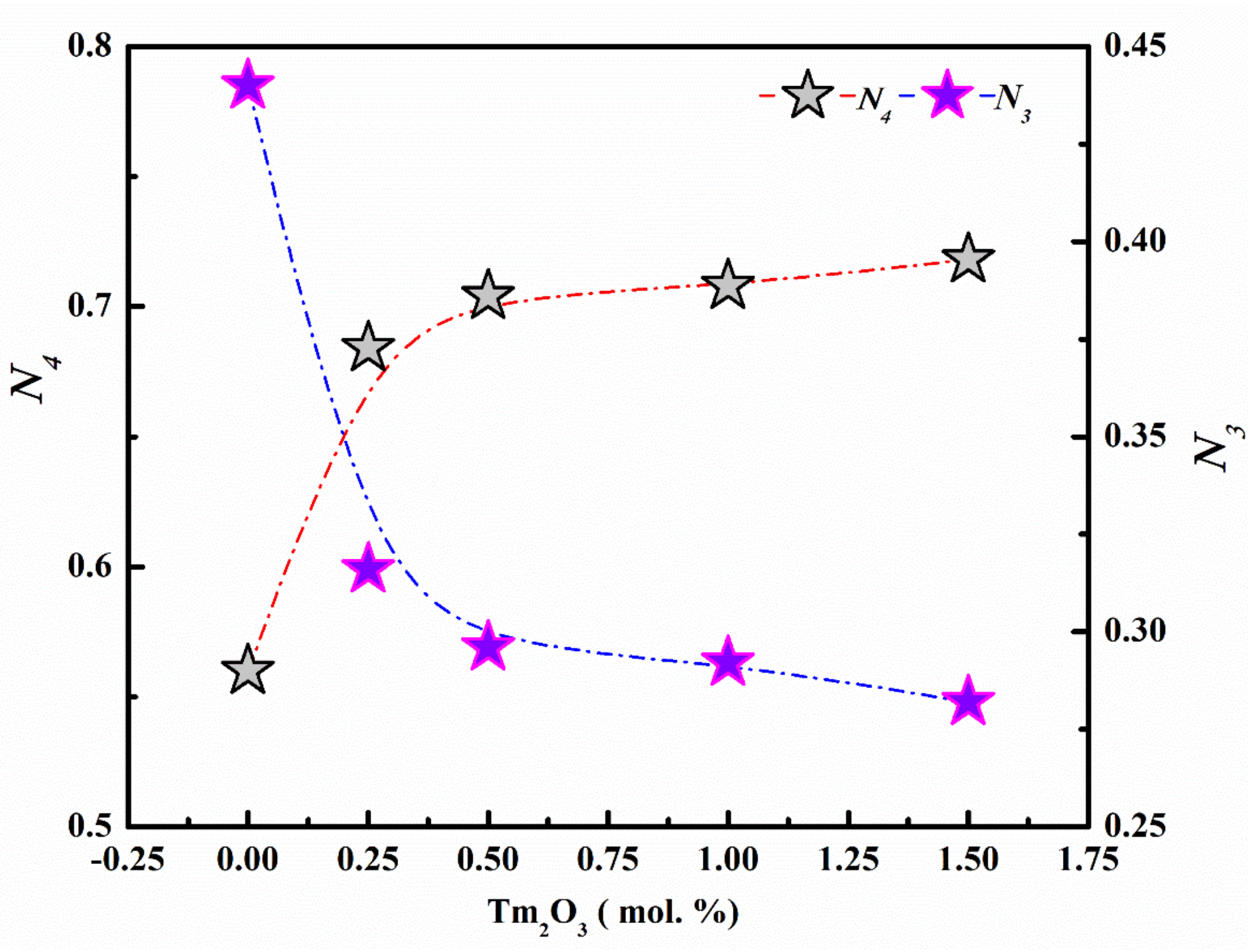

**Fig. 4:** $N_4$& $N_3$ of TGMB-Tm glasses.

**Table 2:** Deconvoluted parameters for pure and $Tm_2O_3$ ions mol% doped glass system.

| | | | | | | | | | | BO4 | BO3 | N4 | N3 |
|---|---|---|---|---|---|---|---|---|---|---|---|---|---|
| TGMB-Tm0 | **C** | - | 721 | 860 | 954 | 1070 | 1223 | 1363 | - | | | | |
| | **A** | - | 19.05 | 5.45 | 23.18 | 16.69 | 20.83 | 14.79 | - | 45.32 | 35.63 | 0.56 | 0.44 |
| TGMBTm0.25 | **C** | 606 | 717 | 859 | 946 | 1073 | 1210 | 1322 | - | 47.06 | 21.75 | 0.684 | 0.316 |
| | **A** | 1.46 | 30.73 | 5.57 | 28.20 | 13.28 | 6.10 | 15.66 | - | | | | |
| TGMB-Tm0.5 | **C** | - | 717 | 856 | 933 | 1064 | 1213 | 1322 | - | 48.26 | 20.28 | 0.704 | 0.296 |
| | **A** | - | 31.46 | 5.15 | 22.72 | 20.39 | 6.34 | 13.94 | - | | | | |
| TGMB-Tm1 | **C** | - | 719 | 860 | 944 | 1065 | 1215 | 1383 | 1511 | 49.88 | 20.52 | 0.708 | 0.292 |
| | **A** | - | 22.71 | 4.55 | 22.62 | 22.70 | 10.14 | 10.38 | 6.88 | | | | |
| TGMB-Tm1.5 | **C** | - | 717 | 858 | 941 | 1070 | 1255 | 1397 | 1513 | 52.71 | 20.70 | 0.718 | 0.282 |
| | **A** | - | 18.35 | 5.19 | 26.33 | 21.20 | 10.79 | 9.91 | 8.24 | | | | |

## 3.3. Optical properties of $Tm^{3+}$-doped germanium–magnesium telluroborate glasses

Germanium–magnesium telluroborate glasses doped with different concentrations of $Tm_2O_3$ were synthesized, and their optical properties were investigated using UV–Vis–NIR absorption and transmission spectroscopy. The optical absorption and transmission spectra of the TGMB-Tm glass samples are shown in Fig. 5(a).

The introduction of $Tm_2O_3$ into the glass composition leads to a noticeable shift in the optical absorption edge. The onset wavelengths were determined to be 332.6, 315.36, 324.65, 324.44, and 325.74 nm for the undoped glass and TGMB-Tm0.25, TGMB-Tm0.5, TGMB-Tm1, and TGMB-Tm1.5 samples, respectively. This variation in the absorption edge reflects changes in the electronic structure of the glass network induced by the incorporation of $Tm^{3+}$ ions.

For the undoped glass and the sample containing 0.25 mol% $Tm_2O_3$, the absorption spectra do not exhibit distinct rare-earth transition bands. In contrast, samples with higher $Tm_2O_3$ concentrations show several characteristic absorption bands centered at approximately 356, 470, 685, 792, 1215, and 1696 nm, which are attributed to the intra-4f electronic transitions of $Tm^{3+}$ ions [18]. The intensity of these absorption bands increases with increasing $Tm_2O_3$ concentration, indicating a higher population of optically active $Tm^{3+}$ centers within the glass matrix.

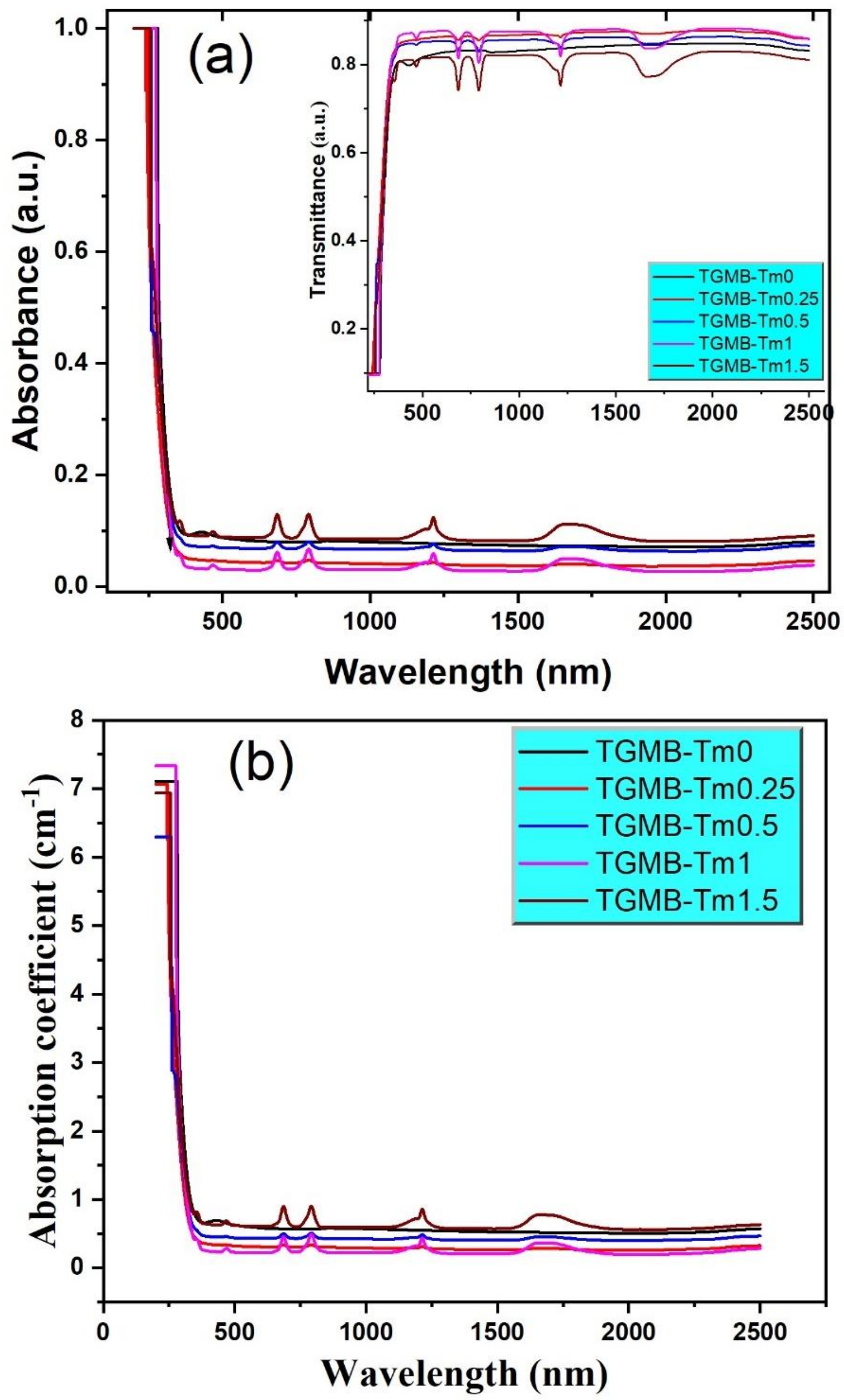

**Fig. 5:** T, A, and α for present TGMB glass-doped thulium ions.

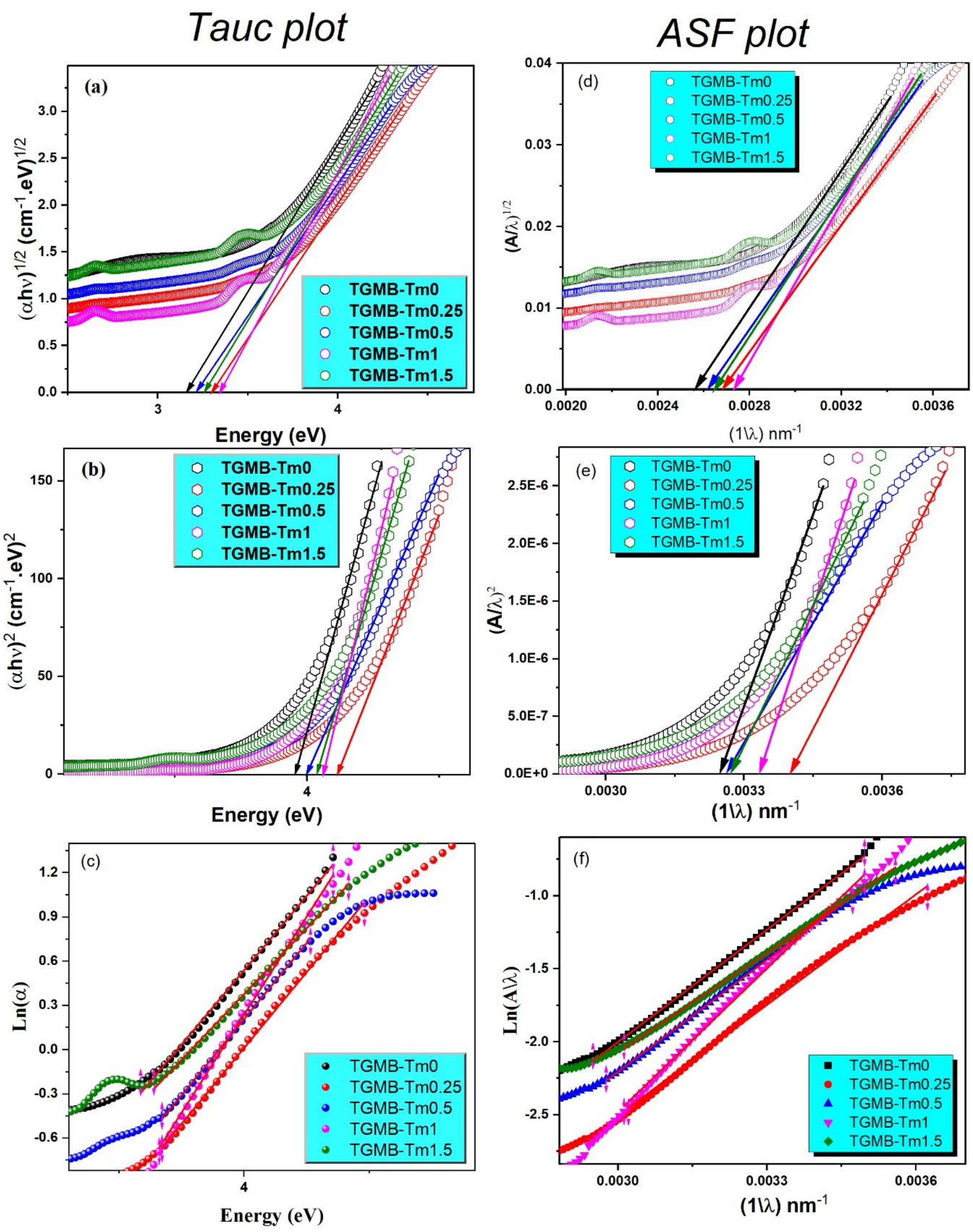

**Fig. 6:** Tauc plot and ASF plot for TGMB glass doped thulium.

## 3.4. Optical transmittance, absorption coefficient, and optical band gap

The transmittance spectra of the TGMB-Tm glass samples are presented in Fig. 5(a). Incorporation of $Tm_2O_3$ generally increases the transmittance of the glasses, except for the sample containing 1.5 mol% $Tm_2O_3$, which exhibits a slight reduction in transmittance.

Figure 5(b) shows the absorption coefficient (α) of the prepared glasses, calculated using the relation:

$$\boldsymbol{\alpha\ (cm^{-1}) = 2.303 \times A / s}$$

where *A* is the measured absorbance and *s* is the glass thickness [19–24]. The results indicate that α decreases with the addition of $Tm_2O_3$ up to 1 mol%, while the TGMB-Tm1.5 sample exhibits the highest absorption coefficient among the studied compositions.

The optical band gap ($E_g$) is a key parameter for evaluating the electronic and optical properties of the glasses. Two approaches were employed to estimate $E_g$: Tauc plot method and the ASF (Absorption Spectrum Fitting) method. In the Tauc approach, the relationship between $(\alpha h\nu)^p$ and photon energy (hν) was analyzed, where $p$ = 0.5, 1.5, 2, or 3, depending on the type of electronic transition [25].

The calculated band gap values are summarized in Table 3. For indirect transitions, $E_g$ values were found to be 3.16, 3.30, 3.22, 3.34, and 3.26 eV for TGMB-Tm0, TGMB-Tm0.25, TGMB-Tm0.5, TGMB-Tm1, and TGMB-Tm1.5, respectively. For direct transitions, $E_g$ values were 3.96, 4.11, 4.01, 4.05, and 4.04 eV, as shown in Fig. 6(a). The band gap values obtained from the ASF method closely agree with those from the Tauc method, with deviations of less than 0.02%, confirming the reliability of the results.

Overall, the results show that the optical band gap exhibits slight variations with $Tm_2O_3$ concentration, reflecting subtle modifications in the electronic structure of the glass network.

| **Samples code** | $E_{g\text{-}indi}$ Tauc | $E_{g\text{-}di}$ Tauc | $E_{g\text{-}indi}$ ASF | $E_{g\text{-}di}$ ASF | $E_U$ | $n_{indi}$ | $n_{di}$ | $\chi^{(1)}$ | $\chi^{(3)}$ $10^{-13}$ | n2 $10^{-12}$ |
|---|---|---|---|---|---|---|---|---|---|---|
| **TGMB-Tm0** | 3.16 | 3.96 | 3.18 | 4.03 | 0.50 | 2.37 | 2.15 | 0.366 | 2.37 | 3.78 |
| **TGMB-Tm0.25** | 3.3 | 4.11 | 3.33 | 4.21 | 0.69 | 2.32 | 2.12 | 0.350 | 2.10 | 3.40 |
| **TGMB-Tm0.5** | 3.22 | 4.01 | 3.25 | 4.05 | 0.70 | 2.35 | 2.14 | 0.359 | 2.25 | 3.61 |
| **TGMB-Tm1** | 3.34 | 4.05 | 3.39 | 4.13 | 0.69 | 2.31 | 2.13 | 0.346 | 2.03 | 3.31 |
| **TGMB-Tm1.5** | 3.26 | 4.04 | 3.29 | 4.05 | 0.70 | 2.34 | 2.13 | 0.355 | 2.17 | 3.51 |

The Urbach energy (E_U) is an important parameter used to characterize the degree of disorder in glass materials. In the present study, E_U for the TGMB-Tm glasses were estimated from the slope of the linear region in a plot of ln(α) versus photon energy (hν), as shown in Fig. 6(a). An inverse relationship between the optical band gap (E_g) and Urbach energy was observed, indicating that an increase in structural disorder corresponds to a decrease in E_g.

The refractive index (n) of the glass samples was estimated using empirical relations based on the measured energy gaps. The refractive index for the indirect transition, $n_0$, was calculated using the relation: $n_0 = (6\sqrt{\frac{5}{E_g}} - 2)$ [32–35]. Additional correlations reported in the literature were also employed [36–38] to extract complementary values of n. The results are presented in Fig. 6(b) and 6(c) and summarized in Table 3.

The results indicate that the refractive index decreases slightly with increasing $Tm_2O_3$ concentration. Specifically, the values for indirect and direct transitions are as follows:

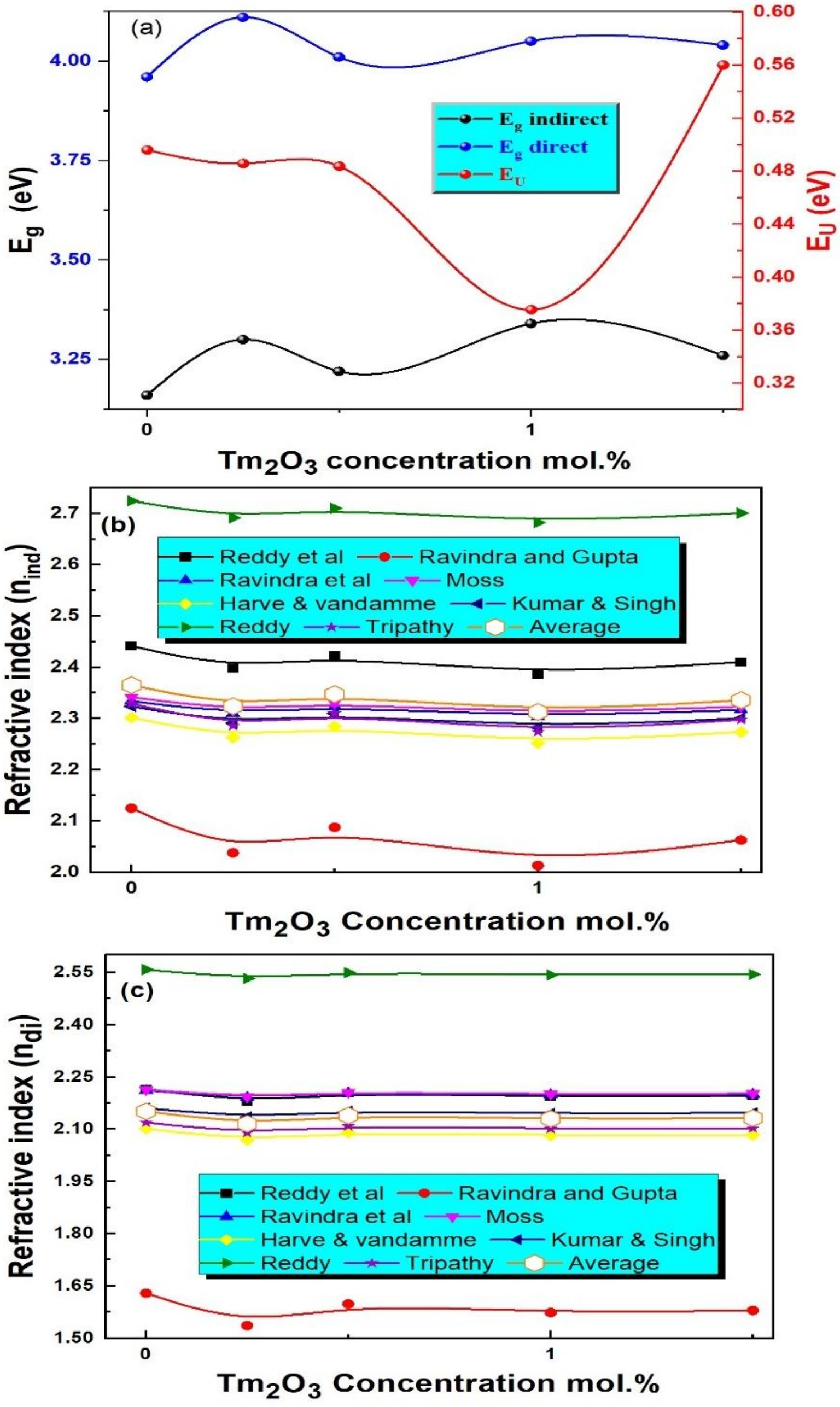

**Fig. 7:** $E_g$, $E_U$, and n for TGMB glasses.

The refractive index (n) was also evaluated as a function of wavelength using the reflectance-based relation:

$n = [(\frac{4R}{(R-1)^2} - k^2]^{1/2} - [(R+1)/(R-1)]$, where $R$ is the reflectance of the sample and $k$ is the extinction coefficient, calculated as: $k = \alpha\lambda/4\pi$ with $\alpha$ being the absorption coefficient and $\lambda$ the wavelength. Figures 8(a) and 8(b) show the variation of n and k with wavelength for the TGMB-Tm glass samples. The results indicate that both n and k depend on the wavelength and the $Tm_2O_3$ content of the glasses. The dielectric constant (ε), derived from the refractive index and extinction coefficient, is plotted versus wavelength in Figures 8(c) and 8(d). These results demonstrate that the optical response of the glasses is influenced by both the composition and the measurement wavelength, consistent with the variations observed in the optical band gap and absorption data.

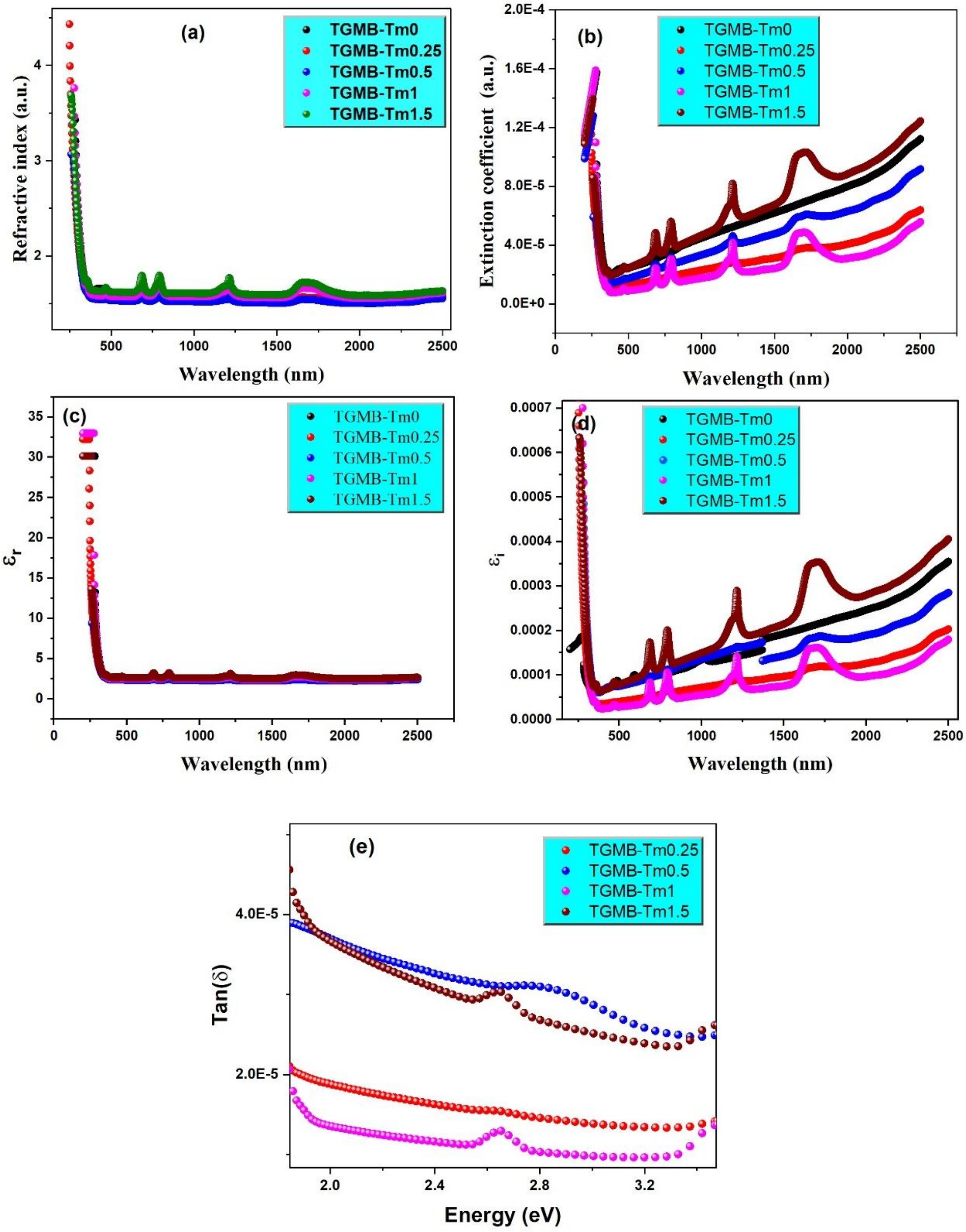

**Fig. 8:** The n, k, $\varepsilon_r$, $\varepsilon_i$, and tan δ of the present fabricated glasses.

The real ($\varepsilon_r$) and imaginary ($\varepsilon_i$) parts of the dielectric constant were found to depend both on the $Tm_2O_3$ content and on the wavelength of incident light. Both $\varepsilon_r$ and $\varepsilon_i$ exhibit trends similar to those observed for the refractive index (n) and extinction coefficient (k), as shown in Figures 8(c) and 8(d). ε related to n and k where, $\varepsilon_r = n^2 - k^2$, $\varepsilon_i = 2nk$.

The energy dissipation of incident photons in the glass system was evaluated in terms of the loss tangent, **tan δ**, which is plotted as a function of wavelength in Fig. 8(e). The results indicate that photon energy absorption and dielectric loss are influenced by both the composition and the wavelength, consistent with the observed optical and electronic behavior of the TGMB-Tm glasses.

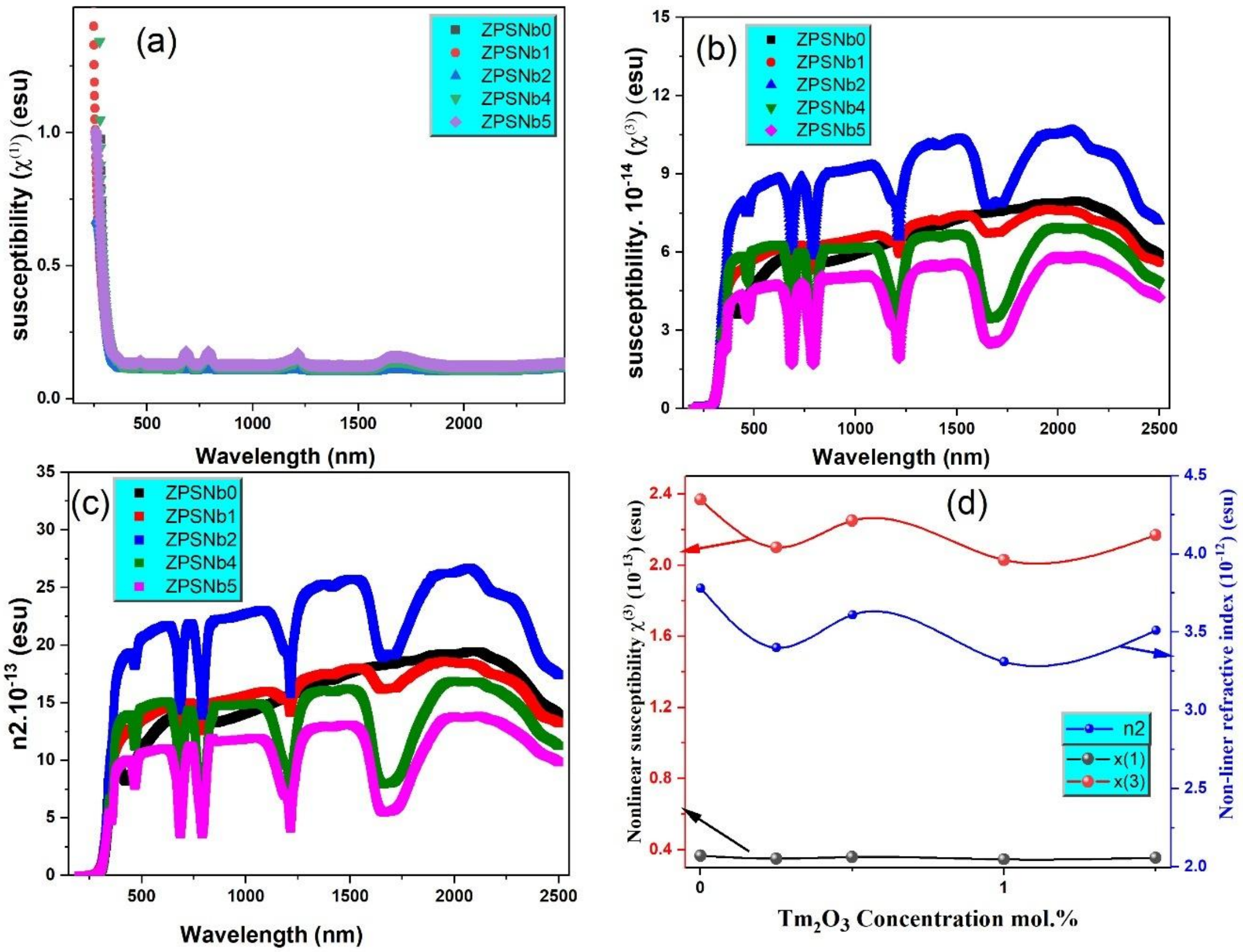

**Fig. 9:** first and third susceptibilities and non-linear index for TGMB glasses.

The nonlinear optical (NLO) properties of the TGMB-Tm glasses were investigated in terms of the first- and second-order susceptibilities and the nonlinear refractive index ($n_2$), as functions of both $Tm_2O_3$ content and wavelength [25]. Figures 9(a), 9(b), and 9(c) show these parameters versus wavelength, while Fig. 9(d) presents their dependence on the $Tm^{3+}$ concentration in the glass system. The results clearly indicate that both the NLO parameters and the nonlinear refractive index are influenced by the addition of $Tm_2O_3$ as well as the wavelength. Specifically, the nonlinear refractive index ($n_2$, in esu) is as follows: TGMB-Tm0= $3.78\times10^{-12}$, TGMB-Tm0.25 = $3.40\times10^{-12}$, TGMB-Tm0.5 = $3.61\times10^{-12}$, TGMB-Tm1 = $3.31\times10^{-12}$, TGMB-Tm1.5 = $3.51\times10^{-12}$ esu. The optical (σ_op) and electrical (σ_el) conductivities were also evaluated using the relations: $\sigma_{Op} = \frac{\alpha c n}{4\pi}$; $\sigma_{el} = \frac{2\lambda\sigma_{Op}}{\alpha}$; where *c* is the speed of light in vacuum, *n* is the refractive index, *α* is the absorption coefficient, and *λ* is the wavelength. Figures 10(a) and 10(b) illustrate the optical and electrical response of the $Tm_2O_3$-doped germanium–magnesium telluroborate glasses, showing high sensitivity to the $Tm_2O_3$ content.

Additionally, the average electronegativity (χ) and metallization (M) of the fabricated glasses were calculated to evaluate the electronic structure. The results, summarized in Table 4, show that both χ and M increase with increasing $Tm_2O_3$ concentration. The measured values are:

χ: 0.397, 0.406, 0.401, 0.409, 0.404

M: 0.849, 0.887, 0.866, 0.898, 0.876

for TGMB-Tm0, TGMB-Tm0.25, TGMB-Tm0.5, TGMB-Tm1, and TGMB-Tm1.5, respectively [3,17,25–28,30].

**Table 4:** Molar Refractivity $R_m$ ($cm^3$/mol), Molar Polarizability $\alpha_m$ ($A^{\circ 3}$), Metallization criterion (M), Reflection loss ($R_L$), Electronegativity (χ), Electronic polarizability($\alpha_o$), Optical basicity ($\wedge_{th}$). Polarizability of cation ($\alpha_{cat}$), Transmission coefficient (T), Steepness (S), and Cohesive energy CE (eV/atom)

| Samples code | $R_m$ | $\alpha_m$ | M | $R_L$ | χ | $\alpha_o$ | $\wedge_{th}$ | $\alpha_{ca}$ | T | S | CE |
|---|---|---|---|---|---|---|---|---|---|---|---|
| **TGMB-Tm0** | 6.32 | 4.86 | 0.397 | 0.603 | 0.849 | 2.74 | 1.628 | 0.102 | 0.717 | 0.0522 | 1.418 |
| **TGMB-Tm0.25** | 6.34 | 4.82 | 0.406 | 0.594 | 0.887 | 2.70 | 1.628 | 0.104 | 0.726 | 0.0374 | 1.445 |
| **TGMB-Tm0.5** | 6.36 | 4.90 | 0.401 | 0.599 | 0.866 | 2.72 | 1.629 | 0.105 | 0.721 | 0.0370 | 1.430 |
| **TGMB-Tm1** | 6.39 | 4.90 | 0.409 | 0.591 | 0.898 | 2.69 | 1.629 | 0.109 | 0.729 | 0.0376 | 1.453 |
| **TGMB-Tm1.5** | 6.43 | 5.01 | 0.404 | 0.596 | 0.876 | 2.71 | 1.630 | 0.113 | 0.724 | 0.0372 | 1.437 |

The average electronic polarizability ($\alpha_0$), molar refractivity ($R_m$, $cm^3/mol$), molar polarizability ($\alpha_m$, $Å^3$), optical basicity ($\Lambda_th$), and cation polarizability ($\alpha_cat$) were evaluated for the TGMB-Tm glasses, and the corresponding values are summarized in Table 4 [15]. The results indicate that $R_m$, $\alpha_m$, $\alpha_cat$, and $\Lambda_th$ increase with increasing $Tm_2O_3$ content in the glass network, whereas $\alpha_0$ shows a slight decrease.

Furthermore, other optical and structural parameters, including reflection loss (RL), transmission coefficient (T), steepness (S), and cohesive energy (CE, eV/atom), were calculated and are also presented in Table 4. These parameters are influenced by the $Tm_2O_3$ content in the glass: RL and S decrease, while T and CE increase. Specifically, RL ranges from 0.603 to 0.591, T from 0.717 to 0.729, S from 0.0522 to 0.0372, and CE from 1.418 to 1.1453 eV/atom.

Figure 11 depicts the variation of $R_m$, $\chi$, $\Lambda_th$, $\alpha_m$, $\alpha_0$, and M as a function of wavelength. The figure demonstrates that these parameters are influenced by both the measurement wavelength and the $Tm_2O_3$ concentration, highlighting the compositional and optical sensitivity of the glass network.

Finally, Table 5 compares the present optical and structural results with previous literature reports. A good degree of agreement is observed for most parameters, confirming the reliability of the current experimental and computational analyses.

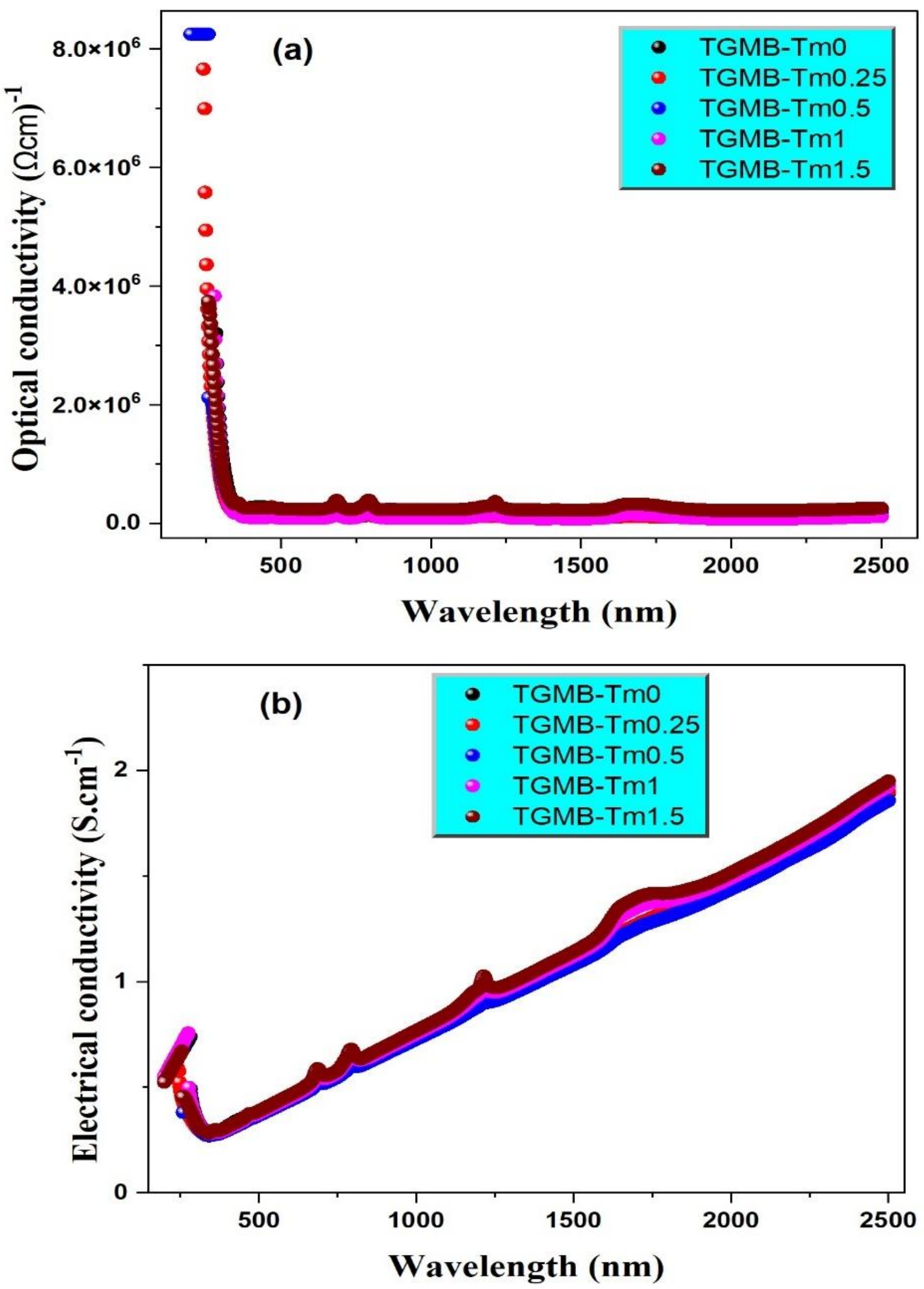

**Fig. 10**: Optical and electrical conductivities versus the wavelength for TGMB glasses.

**Table 5:** Comparison of some structural and optical properties of the present study and previous work.

| Parameters | Present work (TGMB-Tm1.5) | Previous work |
|---|---|---|
| $E_g$ (eV) | 3.29 | 2.93 [33], 2,34 [34], 3.18 [35], 3.57 [36], 3.16 [37] |
| n | 2.34 | 2.416 [33], 1.64 [34], 2.00 [37] |
| Λ | 1.63 | 0.9589 [33]. 1.38 [34], 1.29 [26] |
| $\alpha_{o2-}$ ($A^{o3}$) | 2.71 | 2.4 [35], 2.74 [37] |
| $\alpha_m$ | 5.01 | 11.637 [33], 5.243 [34], 5.83 [36] |
| $\lambda_{cutoff}$ (nm) | | 351.4 [35], |
| M | 0.404 | 0.3828 [33], 0.655 [34], 0.397 [36] |
| $E_U$ | 0.7 | 0.808 [33], 0.435 [34],0.245 [35], 0.31 [37] |

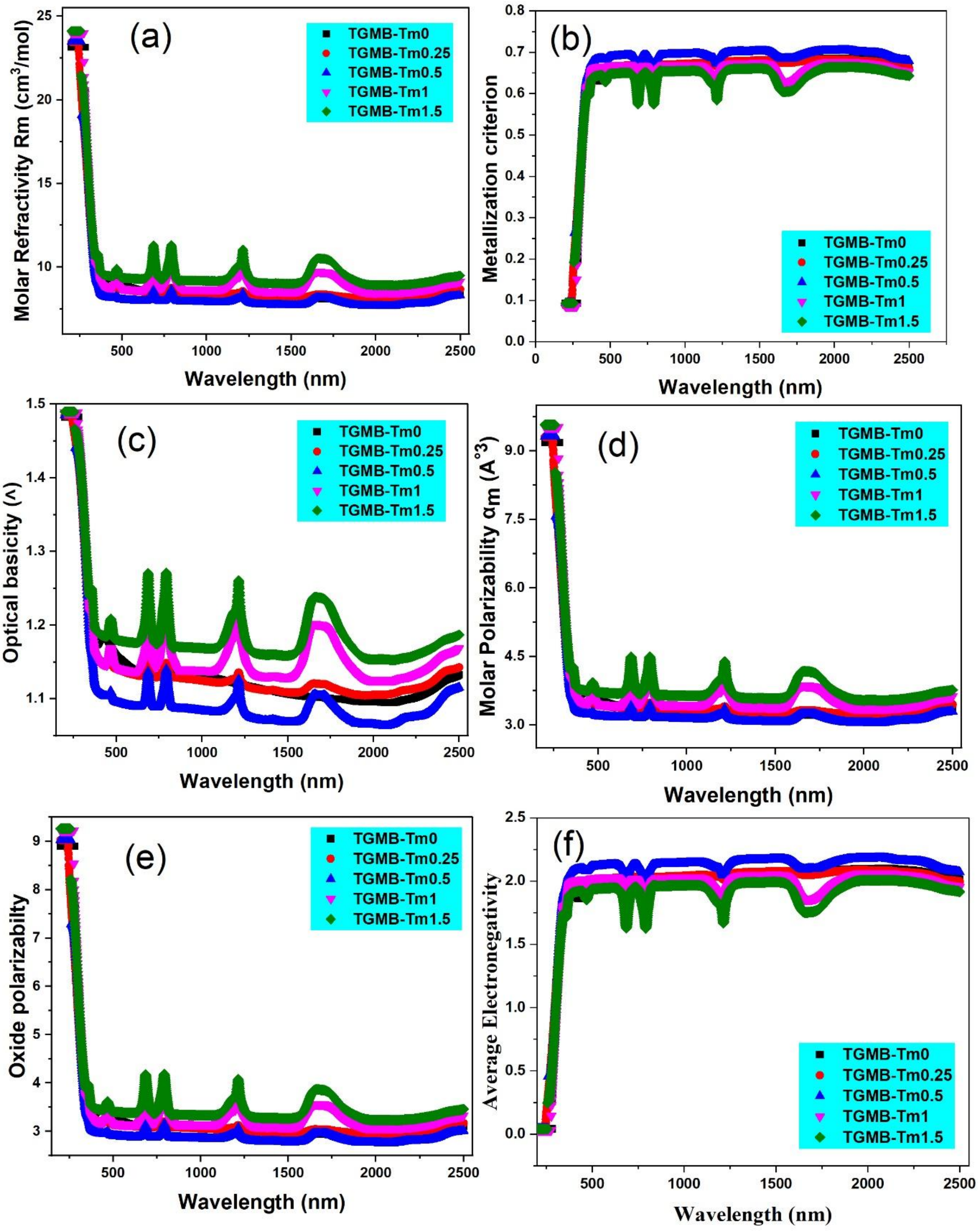

**Fig.11**: $R_m$, $\chi$, $\wedge_{th}$, $\alpha_m$, $\alpha_o$, and M versus the wavelength for TGMB glasses.

### 3.4 Judd Ofelt calculations

According to Judd–Ofelt (J–O) theory, the oscillator strengths, transition probabilities, and line strengths of rare-earth ions can be quantitatively determined from their absorption spectra [38,39]. This theory is based on electric-dipole transitions within the 4f electronic configuration, whose wave functions are effectively shielded by the filled 5s and 5p orbitals. The reduced matrix elements ($U^t$, where t = 2, 4, and 6) required for these calculations were taken from standard tabulated values [40].

The experimental line strength ($S_m$) for each absorption band was calculated using:

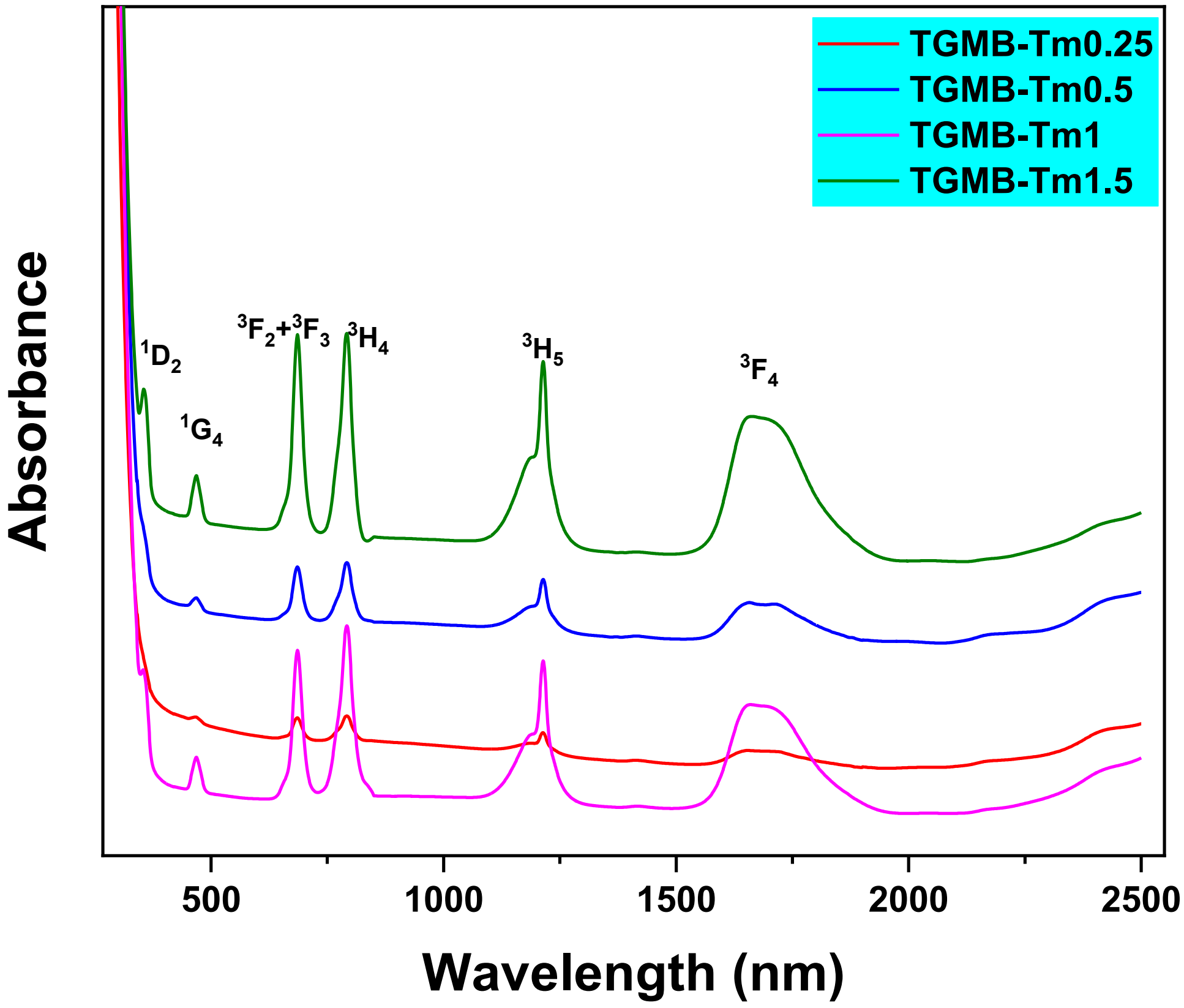

**Fig. 12**: The absorbance of the present glasses as a function of wavelength.

$$S_m = \frac{3c(2J+1)h}{8\pi_3 e^2 \bar{\lambda}} n \left(\frac{3}{n^2+2}\right)^2 \int \sigma(\lambda) d\lambda, \quad (5)$$

where $J$ is the total angular momentum quantum number of the lower level, $\bar{\lambda}$ is the mean wavelength of the absorption band, $n$ is the refractive index of the glass, and $\int\sigma(\lambda)d\lambda$ represents the integrated absorption cross-section. The term $n[3/(n^2 + 2)]^2$ corresponds to the local field correction factor ($\chi_e$d).

The theoretical electric-dipole line strength ($S_e$d) was calculated using:

$$S_{ed} = \sum_{t=2,4,6} \Omega_t |\langle f^n[LSJ]||U^t||f^n[L'S'J']\rangle|^2, \tag{6}$$

where $\Omega_2$, $\Omega_4$, and $\Omega_6$ are the Judd–Ofelt intensity parameters.

Figure 12 shows the absorption spectra of TGMB glasses doped with different concentrations of $Tm^{3+}$ ions (0.25–1.5 mol%). Several well-defined absorption bands corresponding to transitions from the $^3H_6$ ground state to higher-lying excited states are observed at approximately 358, 468, 685, 790, 1214, and 1690 nm. These bands are assigned to the transitions $^1D_2$, $^1G_4$, $^3F_2$ + $^3F_3$, $^3H_4$, $^3H_5$, and $^3F_4$, respectively. The absorption intensity increases systematically with increasing $Tm^{3+}$ concentration, confirming the effective incorporation of thulium ions into the glass network.

The absorption cross-section σ_ab(λ) was calculated using the Beer–Lambert relation [47]:

$$\sigma_{ab}(\lambda) = \frac{2.303 \log(I_o(\lambda)/I(\lambda))}{Nl}$$

(7)

where $I_0$ and I are the incident and transmitted light intensities, respectively, *N* is the concentration of $Tm^{3+}$ ions, and *l* is the sample thickness.

Using the calculated line strengths, the spontaneous radiative transition probability (A_J′J) was evaluated using:

$$A_{J'J} = \frac{64\pi^4 e^2}{3hg\bar{\lambda}^3}\left(\frac{1}{9}n(n^2+2)^2 S_{ed} + n^2 S_{md}\right). \tag{8}$$

where *g* = 2(J′ + 1) is the degeneracy of the upper state, and S_md is the magnetic-dipole line strength, given by:

$$S_{md} = \frac{64\pi^4 e^2}{3h(2J+2)\bar{\lambda}^3} n^3 \left(\frac{h}{2mc}\right)^2 \langle f^n[LSJ]|L + 2S|f^n[L'SJ']\rangle^2, \tag{9}$$

The radiative lifetime (τ_J) of an excited state was calculated from the inverse sum of all radiative transition probabilities:

$$\tau_J = 1/\sum_i A_{J'J}. \tag{10}$$

Table 6 summarizes the absorption band parameters, including integrated cross-sections, full width at half maximum (FWHM), and field correction factors. The $^3F_4 \rightarrow {}^3H_6$ transition near 1690 nm exhibits the highest integrated absorption and the broadest bandwidth, indicating strong near-infrared absorption.

Table 7 compares the experimental (S_exp) and calculated (S_calc) line strengths for different manifolds and $Tm^{3+}$ concentrations. Overall, good agreement between experimental and theoretical values is observed. Minor discrepancies may be attributed to inhomogeneous broadening, local structural disorder, and concentration-dependent interactions within the glass matrix.

The quality of the Judd–Ofelt fitting was evaluated using the root-mean-square deviation:

$$\delta_{rms} = \sqrt{\frac{\sum_{i=1}^{n}(f_i^{exp} - f_i^{calc})^2}{n}} \tag{11}$$

The obtained δ_rms values (0.0133–0.0153 × $10^{-20}$ $cm^2$) indicate a reliable and consistent Judd–Ofelt analysis.

The Judd–Ofelt intensity parameters $\Omega_2$, $\Omega_4$, and $\Omega_6$ for the TGMB-Tm glasses are also listed in Table 7. In most samples, the trend $\Omega_4 < \Omega_6 < \Omega_2$ is observed, suggesting a relatively asymmetric local environment around the $Tm^{3+}$ ions. The comparatively large $\Omega_2$ values indicate significant covalency and structural asymmetry in the Tm–O bonding environment. Variations in the $\Omega_4/\Omega_6$ ratio reflect changes in glass rigidity and vibrational coupling with increasing dopant concentration. The obtained parameters are in good agreement with previously reported values for thulium-doped glass systems [42–45].

Table 8 presents the calculated radiative lifetimes (τ_rad) and branching ratios (β) for various excited states. The $^1D_2$ level exhibits relatively short lifetimes (0.032–0.041 ms), while longer lifetimes are observed for the $^3F_2$ and $^3F_4$ levels, indicating slower radiative decay. The branching ratio analysis reveals that the $^3F_4 \rightarrow {}^3H_6$ transition is dominant, making these glasses promising candidates for near-infrared applications around 2 μm.

Figure 13 illustrates the theoretical emission spectra in the wavelength range 300–2000 nm. Several intense resonance transitions are identified, including $^3F_4 \rightarrow {}^3H_6$, $^3H_4 \rightarrow {}^3H_6$, and $^3P_1 \rightarrow$

$^1D_2$. The increase in line strength with $Tm^{3+}$ concentration confirms enhanced emission probability, although concentration quenching effects may occur at higher dopant levels.

The emission cross-section was calculated using the McCumber relation [53]:

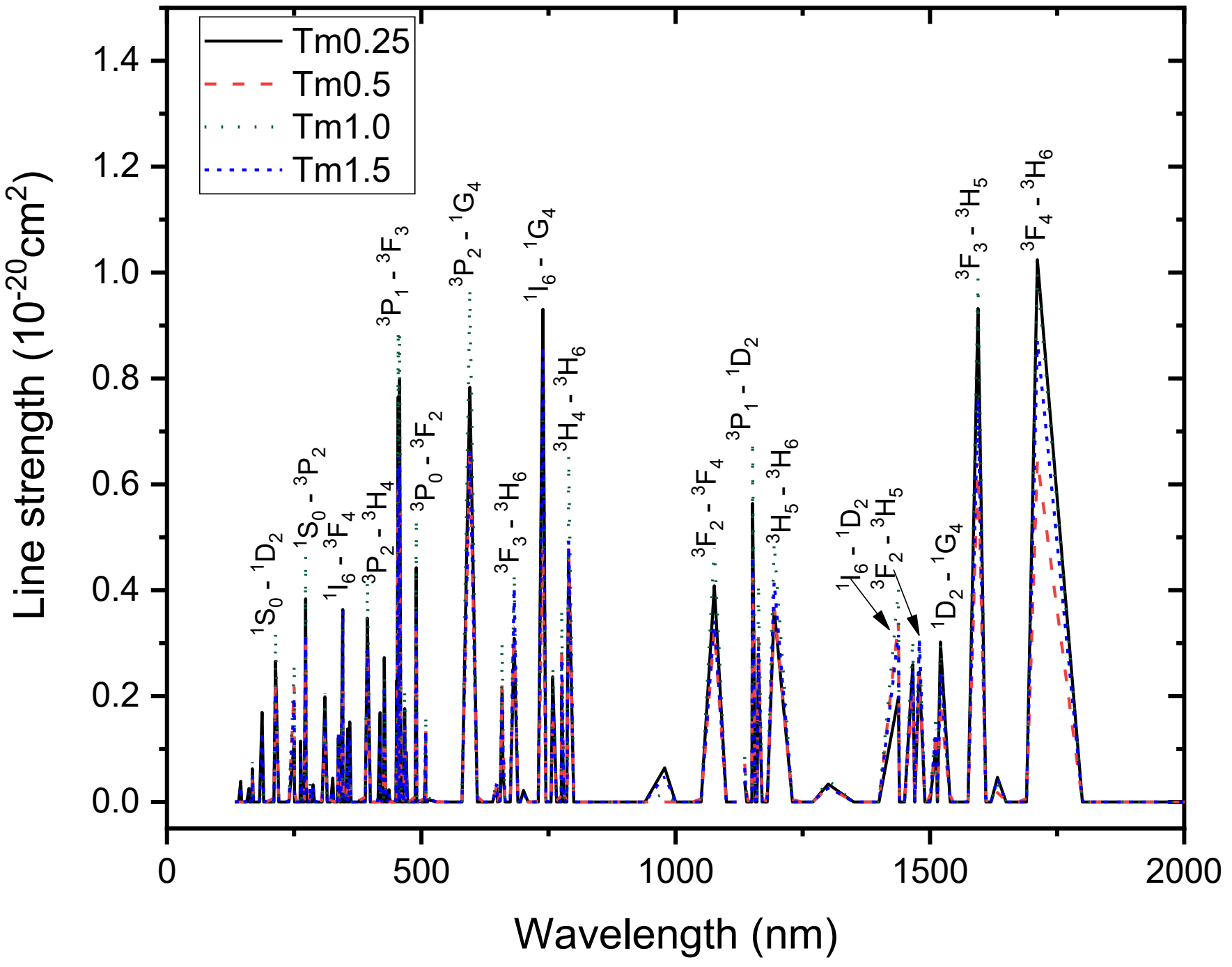

**Fig. 13:** Theoretical spectra of TGMB glass doped with different amounts of $Tm_2O_3$

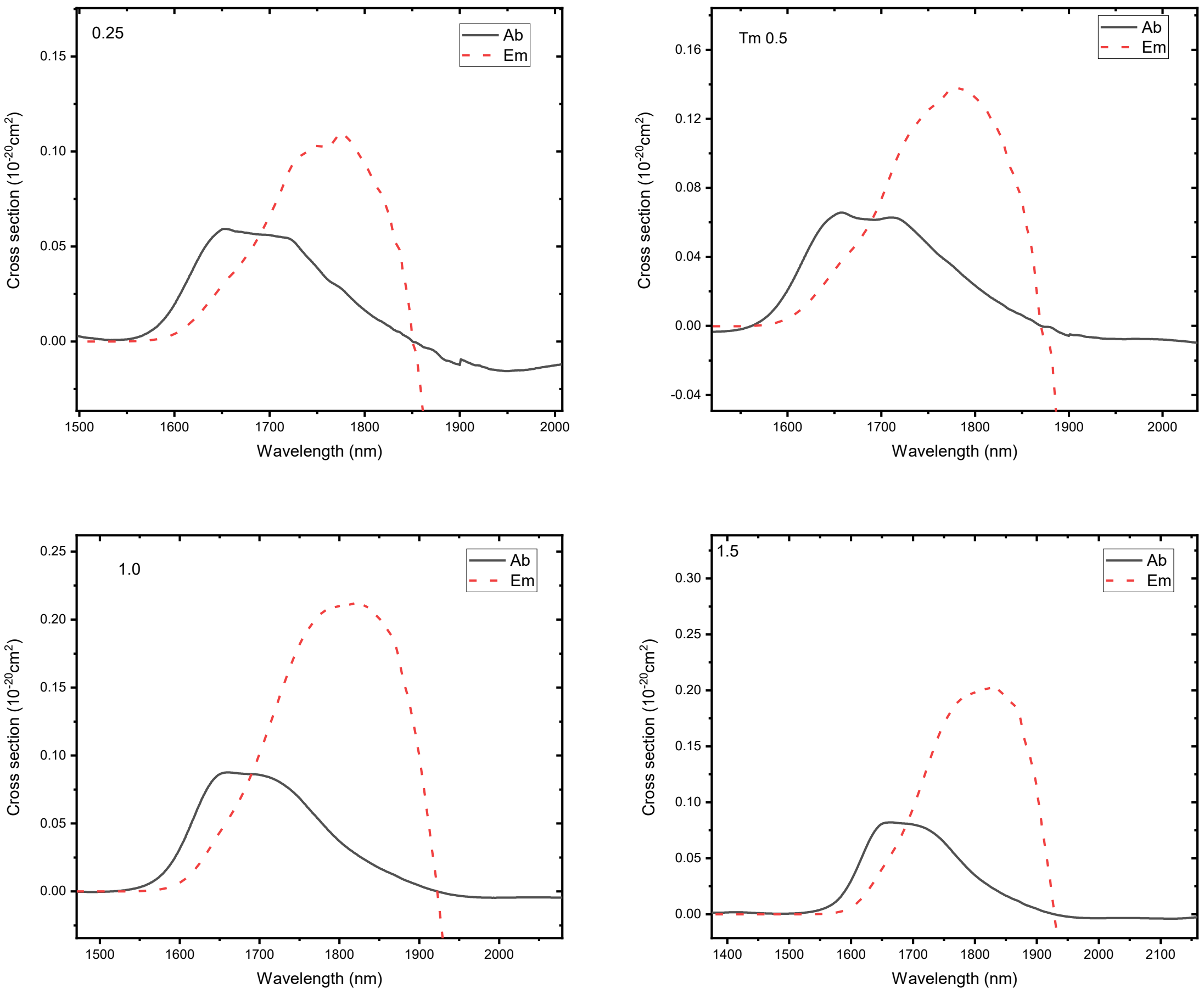

**Fig. 14:** The absorption and emission cross-sections of the spectral line at 1690 nm as a function of wavelength at different content of $Tm_2O_3$.

$$\sigma_{em}(\nu) = \sigma_{ab}(\nu)\exp\left(\frac{\epsilon - h\nu}{kT}\right). \tag{12}$$

Figure 14 compares the absorption and emission cross-sections for all doped samples. The strongest absorption occurs near 1690 nm ($^3F_4 \rightarrow {}^3H_6$), while the emission peak is centered around ~1800 nm, resulting in a Stokes shift of approximately 110 nm. The emission cross-section increases with increasing $Tm^{3+}$ concentration and reaches its maximum at 1.5 mol%, indicating enhanced emission efficiency with a possible onset of concentration quenching.

The dominance of the $^3F_4 \rightarrow {}^3H_6$ transition, combined with its relatively high emission cross-section, favorable branching ratio, and millisecond-scale radiative lifetime, demonstrates efficient near-infrared radiative relaxation. These spectroscopic characteristics, together with the wide transparency window and structural stability of the TGMB glass host, indicate that the present materials are promising candidates for infrared optical amplifiers, mid-infrared light sources, and photonic devices operating near 2 µm.

## Conclusion

The melt-quenching procedure was successfully used to fabricate the present glass system. In the present glass system, the quantity of MgO is replaced by $Tm_2O_3$. The $\rho$ values increase from 3.574 to 4.153 g/cm$^{-1}$, whereas the ($V_m$) values decrease from 21.145 to 19.445 cm$^3$/mol. FT-IR analysis supports the existence and conversion of $BO_3$ and $BO_4$. The conversion of $BO_3$ to $BO_4$ units would lead to a greater BO, which would influence and reinforce the glass network. The optical properties of the current glass samples were examined. The optical energy gap decreased from 3.56 eV to 2.91 eV while the index of refraction increased from 2.25 to 2.44 by increasing the $Tm_2O_3$ content in the glass from 0 mol.% to 8 mol.%. The optical conductivity and optical basicity are also influenced by the ratio of thulium in the specimens. The average electronegativity of the specimens decreased from 0.96 to 0.78 while the average oxide polarization increased from 2.64 to 2.8. The $R_m$ (cm$^3$/mol), and $\alpha_m$ (A$^{\circ 3}$) increased while M and $R_L$ decreased with increased the content of thulium in the present glass. Overall, the properties of germanium-magnesium-telluroborate glass were enhanced by supplying $Tm_2O_3$ in their network. The refractive index shows slight, non-monotonic variations with increasing $Tm_2O_3$ concentration, reflecting the competing effects of structural modification and electronic polarizability. Judd-Ofelt theory has been applied to the present materials. Line strengths, radiative lifetimes, and optical intensity parameters have been computed for the present glass doped with $Tm^{3+}$. The emission and absorption cross-sections of the 1690 nm spectral line have been estimated, as well. The detailed analysis of the present Judd-Ofelt calculations shows that the TGMB glass doped with $Tm^{3+}$ exhibits behavior consistent with desired properties for optical materials, with some transitions being more efficient than others depending on $Tm^{3+}$ concentration. The high branching ratios in certain transitions, alongside the relatively long lifetimes, support its potential use in various photonic technologies. The present glasses doped with $Tm^{3+}$ may find applications in optoelectronics for light amplification and signal processing. $Tm^{3+}$-doped TGMB glass could be forecasted how will behave in practical applications by knowing their branching ratios and lifetimes, which helps with the creation of parts like fiber lasers and amplifiers.

**Acknowledgment**

ChatGPT-4 (OpenAI) was used only for language editing and readability enhancement. No AI tools were used for data generation, analysis, or scientific decision-making.

**Funding**

**Table 5**: Integrated cross-section ($\int \sigma(\lambda)d\lambda$) of the absorption spectrum (band sum), the full width at half maximum (FWHM), and the field correction of REIs in the initial manifold ($\chi_{ed}$) of the present glasses.

| Manifold | λ, nm | 0.25 | | | 0.5 | | | 1.0 | | | 1.5 | | |
|---|---|---|---|---|---|---|---|---|---|---|---|---|---|
| | | $\int \sigma(\lambda)d\lambda$ | FWHM | $\chi_{ed}$ | $\int \sigma(\lambda)d\lambda$ | FWHM | $\chi_{ed}$ | $\int \sigma(\lambda)d\lambda$ | FWHM | $\chi_{ed}$ | $\int \sigma(\lambda)d\lambda$ | FWHM | $\chi_{ed}$ |
| $^1D_2$ | 358 | -- | -- | -- | -- | -- | -- | 0.46591 | 13.57035 | 1.42748 | 0.55391 | 14.50446 | 1.4674 |
| $^1G_4$ | 468 | 0.37804 | 19.40565 | 1.39889 | 0.48402 | 21.39432 | 1.35423 | 0.60339 | 20.00463 | 1.42249 | 0.56074 | 20.33013 | 1.46294 |
| $^3F_2$+$^3F_3$ | 685 | 2.31557 | 24.53744 | 1.39603 | 2.61088 | 25.24556 | 1.35109 | 3.33187 | 26.15789 | 1.41893 | 3.14757 | 26.56177 | 1.45977 |
| $^3H_4$ | 790 | 3.71609 | 35.43724 | 1.39543 | 3.69484 | 33.74566 | 1.35043 | 5.29662 | 37.73912 | 1.41818 | 3.97518 | 33.20237 | 1.4591 |
| $^3H_5$ | 1214 | 4.74168 | 72.68907 | 1.39439 | 4.60496 | 66.02994 | 1.34929 | 6.48784 | 75.69563 | 1.41688 | 5.91292 | 75.47079 | 1.45795 |
| $^3F_4$ | 1690 | 17.86274 | 202.41851 | 1.39403 | 10.83484 | 162.61718 | 1.34888 | 17.51588 | 178.02587 | 1.41643 | 15.81836 | 175.09395 | 1.45754 |

**Table 6**: Experimental and theoretical line strengths ($10^{-20}$ cm$^2$) of TGMB glass doped with $Tm_2O_3$.

| Manifold | λ, nm | $\Delta E$, cm$^{-1}$ | TGMB-Tm0.25 | | TGMB-Tm0.5 | | TGMB-Tm1.0 | | TGMB-Tm1.5 | |
|---|---|---|---|---|---|---|---|---|---|---|
| | | | $S_{exp}$ | $S_{calc}$ | $S_{exp}$ | $S_{calc}$ | $S_{exp}$ | $S_{calc}$ | $S_{exp}$ | $S_{calc}$ |
| $^1D_2$ | 358 | 27933.0 | -- | -- | -- | -- | 0.1234 | 0.1034 | 0.1427 | 0.1445 |
| $^1G_4$ | 468 | 21367.5 | 0.0782 | 0.1007 | 0.1034 | 0.0673 | 0.1227 | 0.101 | 0.1109 | 0.0893 |
| $^3F_2$+$^3F_3$ | 685 | 14598.5 | 0.3278 | 0.3264 | 0.3819 | 0.3754 | 0.4641 | 0.469 | 0.4261 | 0.4326 |
| $^3H_4$ | 790 | 12658.2 | 0.4563 | 0.4541 | 0.4689 | 0.4629 | 0.64 | 0.625 | 0.4669 | 0.4722 |
| $^3H_5$ | 1214 | 8237.2 | 0.3792 | 0.3824 | 0.3806 | 0.396 | 0.5106 | 0.5193 | 0.4523 | 0.4398 |
| $^3F_4$ | 1690 | 5917.2 | 1.0264 | 1.0248 | 0.6434 | 0.6494 | 0.9906 | 0.9974 | 0.8693 | 0.8714 |
| $\delta_{rms}(\times 10^{-20}$ cm$^2)$ | | | 0.0133 | | 0.0314 | | 0.0204 | | 0.0153 | |

**Table 7**: Judd Ofelt intensity parameters of TGMB glass doped with $Tm_2O_3$.

| **Sample** | **$\Omega_2$** | **$\Omega_4$** | **$\Omega_6$** | **$\Omega_4$/ $\Omega_6$** | **Trend** |
|---|---|---|---|---|---|
| TGMB-Tm0.25 | 1.2469± 04 | 0.4225± 0.03 | 0.2097± 0.01 | 5.946 | **$\Omega_6 < \Omega_4 < \Omega_2$** |
| TGMB-Tm0.5 | 1.0315±0.12 | 0.0029±0. 10 | 0.4075±0.03 | 2.531 | **$\Omega_4 < \Omega_6 < \Omega_2$** |
| TGMB-Tm1.0 | 1.5162± 0.06 | 0.1021± 0.05 | 0.4751± 0.02 | 2.433 | **$\Omega_4 < \Omega_6 < \Omega_2$** |
| TGMB-Tm1.5 | 1.0317± 0.05 | 0.3236± 0.04 | 0.3593± 0.02 | 2.871 | **$\Omega_4 < \Omega_6 < \Omega_2$** |
| Fluroborosilicate [48] | 1.376 | 0.838 | 0.239 | 3.506 | **$\Omega_2 > \Omega_4 > \Omega_6$** |
| Fluorophosphate [49] | 2.32 | 1.74 | 1.64 | 1.061 | **$\Omega_6 < \Omega_4 < \Omega_2$** |
| Borate [50] | 4.37 | 0.05 | 1.41 | 0.035 | **$\Omega_4 < \Omega_6 < \Omega_2$** |
| Tellurite [51] | 4.78 | 1.71 | 1.20 | 1.43 | **$\Omega_6 < \Omega_4 < \Omega_2$** |

**Table 8.** Calculated values of radiative lifetime (in ms) and branching ratio for TGMB glass containing $Tm^{3+}$-ions.

| Transition | λ | TGMB- | | TGMB- | | TGMB- | | TGMB- | | ($\tau_{rad}$) Other values |
|---|---|---|---|---|---|---|---|---|---|---|
| | | $\tau_{rad}$ | β% | $\tau_{rad}$ | β% | $\tau_{rad}$ | β% | $\tau_{rad}$ | β% | |
| $^1D_2 \rightarrow {^1G_4}$ | 1520.7 | 0.035 | 0.63 | 0.041 | 0.61 | 0.033 | 0.62 | 0.032 | 0.6 | 0.034 [47], 0.02 [48], 0.092 [6] |
| $^3F_2$ | 787.4 | | 3.74 | | 2.37 | | 2.5 | | 3.57 | |
| $^3F_3$ | 758.2 | | 4.87 | | 5.71 | | 5.24 | | 4.84 | |
| $^3H_4$ | 658.9 | | 5.47 | | 8.8 | | 7.85 | | 6.55 | |
| $^3H_5$ | 513.9 | | 0.23 | | 0.55 | | 0.44 | | 0.41 | |
| $^3F_4$ | 454.7 | | 60.36 | | 72.4 | | 70.9 | | 58.46 | |
| $^3H_6$ | 359.2 | | 24.7 | | 9.51 | | 12.4 | | 25.57 | |
| $^1G_4 \rightarrow {^3F_2}$ | 1632.9 | 0.750 | 0.51 | 0.843 | 0.27 | 0.663 | 0.33 | 0.666 | 0.48 | 0.23 [46], 0.458 [47], 0. 501 [48], 0.22 [51], 1.24 [6] |
| $^3F_3$ | 1512.2 | | 1.84 | | 2.38 | | 2.21 | | 2.29 | |
| $^3H_4$ | 1162.8 | | 12.1 | | 14.2 | | 13.5 | | 12.52 | |
| $^3H_5$ | 776.2 | | 35.85 | | 48.2 | | 43.4 | | 41.62 | |
| $^3F_4$ | 648.6 | | 6.08 | | 7.53 | | 7.08 | | 7.38 | |
| $^3H_6$ | 470.3 | | 43.61 | | 27.3 | | 33.4 | | 35.71 | |
| $^3F_2 \rightarrow {^3F_3}$ | 20449. | 1.001 | 0 | 1.037 | 0 | 0.833 | 0 | 0.855 | 0 | 0.35 [46], 0.762 [47], 0. 726 [52], 0.61 [51], 1.88 [6], 0.99 [49] |
| $^3H_4$ | 4038.8 | | 1.16 | | 0.73 | | 0.84 | | 0.85 | |
| $^3H_5$ | 1479.3 | | 12.72 | | 10.4 | | 10.4 | | 13.5 | |
| $^3F_4$ | 1076 | | 54.62 | | 37.0 | | 42.1 | | 39.45 | |
| $^3H_6$ | 660.6 | | 31.49 | | 51.7 | | 46.6 | | 46.2 | |
| $^3F_3 \rightarrow {^3H_4}$ | 5032.7 | 0.749 | 0.22 | 0.804 | 0.15 | 0.646 | 0.16 | 0.625 | 0.18 | 0.23 [46], 0.496 [47], 0.23 [51], 1.118 [6], 0.58 [49] |
| $^3H_5$ | 1594.6 | | 16.89 | | 11.3 | | 13.6 | | 11.51 | |
| $^3F_4$ | 1135.7 | | 10.85 | | 12.0 | | 9.97 | | 10.02 | |
| $^3H_6$ | 682.7 | | 72.03 | | 76.5 | | 76.2 | | 78.29 | |
| $^3H_4 \rightarrow {^3H_5}$ | 2334.3 | 0.986 | 3.43 | 1.071 | 2.00 | 0.859 | 1.78 | 0.871 | 2.95 | 0.36 [46], 0. 818 [47], 0. 834 [38] 0.885 [52], 0.35 [51], 0.23 [39] 2.179 [6], |
| $^3F_4$ | 1466.7 | | 11.94 | | 10.6 | | 9.89 | | 11.15 | |
| $^3H_6$ | 789.8 | | 84.63 | | 87.3 | | 88.3 | | 85.91 | |
| $^3H_5 \rightarrow {^3F_4}$ | 3946.3 | 3.077 | 0.67 | 3.341 | 0.79 | 2.808 | 0.88 | 2.698 | 0.78 | 1.62 [46], 4.044 [47], 4.356 [52], 1.97 [51], 1.17 [50], 11.66 |
| $^3H_6$ | 1193.7 | | 99.33 | | 99.21 | | 99.12 | | 99.22 | |
| $^3F_4 \rightarrow {^3H_6}$ | 1711.4 | 3.851 | 100 | 4.555 | 100 | 3.72 | 100 | 3.561 | 100 | 2.26 [46], 4.10 [47], 4.81 [52], 5.01 [48] |